\documentclass[a4paper,11pt]{article}
\usepackage{pos}
\usepackage{mathtools}

\newcommand{\eqn}[1]{(\ref{#1})}
\newcommand{\be}{\begin{equation}}
\newcommand{\ee}{\end{equation}}
\newcommand{\bea}{\begin{eqnarray}}
\newcommand{\eea}{\end{eqnarray}}
\allowdisplaybreaks

\title{$\varrho$-Poincar\'e: bicrossproduct structure, $\star$-products and quantum Lie algebra}

\author*[a]{Luca Scala}

\affiliation[a]{Faculty of Physics and Astronomy, University of Wrocław,\\
Maksa Borna 9, 50-204 Wrocław, Poland}

\emailAdd{luca.scala@uwr.edu.pl}

\abstract{We discuss the bicrossproduct structure of the quantum group $\varrho$-Poincar\'e and of the dual quantum universal enveloping algebra, expanding the construction to general Lie algebra-type deformations of Poincar\'e coming from classical $r$-matrices. We review the relation between different bases of the quantum universal enveloping algebra of $\varrho$-Poincar\'e and noncommutative $\star$-products defined on the $\varrho$-Minkowski spacetime, analysing some of their relevant features. Furthermore, we comment on the role of physical bases and introduce the $\varrho$-Poincar\'e quantum Lie algebra.}

\FullConference{%
  Workshop on Noncommutative and Generalised Geometry in String theory, Gauge theory and Related Physical Models\\
  18-25 September 2023\\
  Mon-Repos, Corfu, Greece
}

\tableofcontents

\begin{document}
\maketitle

\section{Introduction}\label{sec:intro}
The quest for a theory of quantum gravity is far away from an end. During the last decades a plethora of possible candidates to assume this role were proposed, ranging from exact theories to effective ones. A vast number of them deal, at least effectively, with deformations of classical structures such as spacetimes and their isometry groups. Noncommutative spacetimes (NCST) and the closely related theory of quantum group (QG) deformations of their symmetry groups and algebras fall into this categorization, providing, furthermore, theoretical models for a wide range of disciplines as well as excellent auxiliary tools for other modern theories.

In this context, the most studied noncommutative deformations of the Minkowski spacetimes are of the form
\begin{equation}\label{B.1}
[x^\mu,x^\nu]=i\alpha^2 \Theta^{\mu \nu}(\alpha^{-1}x),
\end{equation}
where $x^\mu$ are self-adjoint operators whose expectation values give the spacetime events, $\alpha$ is a length parameter and $\Theta^{\mu \nu}$ a constant matrix that can be expanded as
\begin{equation}\label{B.2}
\Theta^{\mu \nu}(\alpha^{-1}x)={\Theta^{\mu \nu}}^{(0)}+\alpha^{-1}{{\Theta^{\mu \nu }}_\rho}^{(1)}x^\rho+\dots
\end{equation}
Usually, this expansion is truncated to the first order due to heuristic quantum gravity arguments. Since $\alpha$ is a length parameter, often identified with the Planck length, and in the ordinary physical regimes (i.e., far away from the Planck length) we expect quantum gravity effects (i.e., spacetime noncommutativity) to be irrelevant, we require the right hand-side of~\eqref{B.1} to be linear or quadratic in $\alpha$\footnote{To be precise, it is in principle possible to consider zero order terms in $\alpha$, but these terms bring small corrections compared to the first two orders of the expansion.}; in this way, operating the ``quasiclassical limit'' (sending $\alpha\rightarrow 0$), we recover the classical commutative Minkowski spacetime.

It is clear that a relation like~\eqref{B.1} will not preserve, in general, Poincar\'e invariance. However, since $\Theta^{\mu \nu}(\alpha^{-1}x)$ is a constant matrix and not a tensor, it is possible to keep Poincar\'e-like symmetries in a deformed fashion. These new covariance groups for NCSTs are called quantum Poincar\'e groups, and the respective deformed algebras of symmetry, quantum universal enveloping Poincar\'e algebras.\footnote{Note that actually a more precise notion of quantum Poincar\'e Lie algebras was proposed. See Section~\ref{sec:qa} for further details.}

The class of spacetimes obtained setting ${{\Theta^{\mu \nu }}_\rho}^{(1)}=0$ in the first order expansion \eqref{B.2}, is usually called ``canonical'' (or ``Moyal'', or ``$\theta$'') Minkowski, while setting ${\Theta^{\mu \nu}}^{(0)}=0$ one obtains the so-called ``Lie algebra-type'' deformations. A general, yet no complete\footnote{The classification is complete for $r$-matrices satisfying a classical Yang-Baxter equation, but not for ones satisfying a modified one.}, classification of these possible deformations in 3+1 dimensions was worked out in~\cite{zakrzewski} in the context of Poincar\'e classical $r$-matrices.

In this work we will consider a particular Lie algebra-type noncommutative spacetime, called $\varrho$-Minkowski, that we will define in Section \ref{sec:rhospace}. In Section \ref{sec:QG} we recall the structure of its quantum group of symmetries and the definition of a bicrossproduct Hopf algebra, showing in the end that this quantum group has a bicrossproduct structure. Furthermore, expanding the analysis carried out in \cite{Fabiano:2023uhg}, we generalise the construction to arbitrary Lie algebra-type quantum groups, showing their explicit bicrossproduct nature. In Section \ref{sec:QUEA} we turn to the dual picture, that of the quantum universal enveloping algebra, writing the bicrossproduct basis for the $\varrho$-deformed case. In Section \ref{sec:star} we review the $\star$-products associated to the two different bases introduced for the quantum universal enveloping algebra and list some of their features. Then, in Section \ref{sec:qa}, we make some considerations about the physical choice of basis of the quantum universal enveloping algebra and argue that the right structure to consider might be, instead, the quantum Lie algebra of $\varrho$-Poincaré, that we introduce here. Section \ref{sec:concl} deals with the conclusions of our analysis.

\section{The \texorpdfstring{$\varrho$}{}-Minkowski spacetime}\label{sec:rhospace}
In this work we will focus on the QG\footnote{Although in literature the term quantum group is used to refer to both the quantised algebras of functions over a group and the quantum universal enveloping algebras, we will call by it only the former.} and the quantum universal enveloping algebra (QUEA) associated with a particular noncommutative spacetime, namely, $\varrho$-Minkowski. In 3+1 dimensions this spacetime, denoted as $\mathcal{M}_\varrho$, can be realized in terms of the following commutation relations
\begin{equation}\label{algebra}
\begin{aligned}
\relax[x^1,x^0]&=i\varrho x^2\,,\\
[x^2,x^0]&=-i\varrho x^1 \,,
\end{aligned}
\end{equation}
all other commutators being zero.

This form of noncommutativity is a special subclass of the one found in~\cite{gutt} via a $\star$-product approach and of the one defined in~\cite{Lukierski} via Drinfel'd twist deformations of the Poincar\'e universal enveloping algebra. Recently, in~\cite{Mercati:2023apu}, a broad class of deformations of Minkowski spacetimes, all coming from the classification of the Poincar\'e $r$-matrices worked out in~\cite{zakrzewski}, was called by the name ``T-Minkowski'', and it was proven there that these deformations are compatible with braiding prescriptions to obtain multipoint algebras. $\varrho$-Minkowski is one of them.

Physical application of this kind of noncommutativity were studied, for instance, in~\cite{Ciric:2017rnf, dimi1, dimi2}, in connection with quasinormal modes of
Reissner–Nordström black holes, and some phenomenological considerations related to relative locality were discussed in~\cite{Amelino-Camelia:2011ycm}. Field theories defined on $\varrho$-Minkowski were considered in~\cite{DimitrijevicCiric:2018blz, Hersent:2023lqm} and, in the semiclassical context,  in~\cite{Kurkov:2021kxa}. 

In Lie algebra-type noncommutative Minkowski spacetimes a sort of ``quantum myopia'' effect arises, related to the fact that observers can only sharply localize states in the origin of their reference frames and not in general translated and Lorentz-rotated ones. Quantum features of observers and localization induced by the $\varrho$-deformation were studied in~\cite{LSV}. It is also worth to note that this model displays an angular-type noncommutativity, as extensively discussed in~\cite{Lizzi:2021dud}. In that paper the spectrum of the time operator was also analysed, and it was found to be discrete.

\section{The \texorpdfstring{$\varrho$}{}-Poincar\'e QG}\label{sec:QG}
We define the $\varrho$-Poincar\'e quantum group $\mathcal{C}_\varrho(P)$ as the Hopf-algebra deformation of the algebra of continuous functions on the Poincar\'e group manifold under which the commutation relations \eqref{algebra} are covariant. In this sense it is the quantum group of isometries of $\varrho$-Minkowski.

The standard procedure employed to obtain this deformation relies on the following classical $r$-matrix:
\begin{equation}\label{rrho}
    r=i\varrho (P_0 \wedge M_{12}),
\end{equation}
where $P_\mu$ and $M_{\mu\nu}$ are the translation and Lorentz generators of the Poincar\'e algebra $\mathfrak{p}$. For a review on this construction see, for instance,~\cite{charipressley, ZakrzewskiInventsKPGroup, Lukierski_kappaPoincareanydimension}.

Evaluating the Schouten brackets $[[r,r]]$, it is straightforward to check that this $r$-matrix satisfies the classical Yang-Baxter equation (CYBE), and therefore there exists an admissible Drinfel'd twist operator that allows for a deformation of the dual universal enveloping algebra, as we will show in the following section.

The full structure of the quantum group is given by the following structure maps~\cite{Lizzi:2021dud}: 
\begin{align}\label{Crho}
\left[ a^\mu, a^\nu \right] &=i \varrho [{\delta^\nu}_0 (a^2 {\delta^\mu}_1-a^1 {\delta^\mu}_2) -{\delta^\mu}_0 (a^2 {\delta^\nu}_1-a^1 {\delta^\nu}_2)]\,,\nonumber 
\\
\left[ {\Lambda^\mu}_\nu, {\Lambda^\varrho}_\sigma \right] &=0\,, \nonumber 
\\
\left[ {\Lambda^\mu}_\nu, a^\rho \right] &=i \varrho \left[{\Lambda^\varrho}_0 ({\Lambda^\mu}_1 g_{2\nu} -{\Lambda^\mu}_2 g_{1\nu} )-{\delta^\rho}_0 (\Lambda_{2\nu} {\delta^\mu}_1 -\Lambda_{1\nu} {\delta^\mu}_2) \right]\,, \nonumber 
\\
\Delta ({\Lambda^\mu}_\nu)&= {\Lambda^\mu}_\alpha \otimes {\Lambda^\alpha}_\nu\,, \nonumber 
\\
S({\Lambda^\mu}_\nu) &={(\Lambda^{-1})^\mu}_\nu\,, 
\\
\varepsilon ({\Lambda^\mu}_\nu)&={\delta^\mu}_\nu\,, \nonumber 
\\
\Delta (a^\mu)&={\Lambda^\mu}_\nu \otimes a^\nu+ a^\mu \otimes 1\,,  \nonumber 
\\
S(a^\mu) &=-a^\nu {(\Lambda^{-1})^\mu}_\nu\,,  \nonumber 
\\
\varepsilon (a^\mu)&=0\,,\nonumber 
\end{align}
where $\Lambda^\mu{}_\nu$ and $a^\mu$ are self-adjoint operators whose expectation values give the Lorentz and translational parameters of the Poincar\'e group $P$, $\Delta$ is the coproduct, $S$ the antipode and $\epsilon$ the counit.

It is, then, possible to check explicitly that $\mathcal{M}_\varrho$ is covariant under the left $\varrho$-Poincar\'e coaction
\begin{equation}
    x^\mu\in \mathcal{M}_\varrho\rightarrow x^{\mu '}=\Lambda^\mu{}_\nu \otimes x^\nu +a^\mu \otimes 1\in \mathcal{C}_\varrho(P)\otimes \mathcal{M}_\varrho.
\end{equation}

\subsection{Bicrossproduct Hopf algebras}\label{sec:bicrossHopf}
A bicrossproduct Hopf algebra is a particular structure that allows to factorize an Hopf algebra as a tensor product of Hopf algebras acting and coacting one onto the other. The reason for us to be interested in this construction is that it represent a natural way to lift the notion of semidirect products of groups to the Hopf algebra context. In particular, a well known feature of the Poincar\'e group is that it can be viewed as a semidirect product of the Lorentz group $SO(1,3)$ and the translational sector $T_4$, i.e.,
\begin{equation}
    P=ISO(1,3)=T_4\rtimes SO(1,3).
\end{equation}
The interesting point, here, is that $T_4$ is isomorphic to the Minkowski spacetime $\mathcal{M}$, and this in turn means that we can obtain the latter as an homogeneous space
\begin{equation}
    \mathcal{M}\cong ISO(1,3)/SO(1,3).
\end{equation}

In the previous subsection we introduced $\mathcal{C}_\varrho(P)$ as the Hopf algebra of quantum isometries of $\mathcal{M}_\varrho$. We would like to understand if there is a way to proceed in the opposite direction, defining $\mathcal{M}_\varrho$ as a sort of ``quantum homogeneous space'' arising from $\mathcal{C}_\varrho(P)$. We will show that the bicrossproduct structure of this quantum group allows for this construction.

In order to do so we define, here, a bicrossproduct structure following~\cite{MAJID1990, Majid_1995}.
Let $\mathcal{X},\mathcal{A}$ be two Hopf algebras; a \textit{bicrossproduct Hopf algebra} $\mathcal{X} \vartriangleright \! \blacktriangleleft \mathcal{A}$ is the tensor product $\mathcal{X} \otimes \mathcal{A}$ endowed with two additional structure maps:
\begin{equation}
\begin{aligned}
\triangleleft &: \mathcal{A}\otimes \mathcal{X}\rightarrow \mathcal{A} \hspace{0.5cm} \text{(right action)},\\
\beta &: \mathcal{X}\rightarrow \mathcal{A}\otimes \mathcal{X} \hspace{0.5cm} \text{(left coaction)},
\end{aligned}    
\end{equation}
such that
\begin{equation}\label{coac}
    \begin{aligned}
        a\triangleleft (xy) &=(a\triangleleft x)\triangleleft y,\\
        1\triangleleft x &=\varepsilon(x)1,\\
        (a\cdot b)\triangleleft x&=(a\triangleleft x_{(1)})(b\triangleleft x_{(2)}),\\
        (id \otimes \beta)\circ \beta &=(\Delta \otimes id) \circ \beta,\\
        (\varepsilon \otimes id) \circ \beta &= id,\\
        \beta(1) &= 1\otimes 1,
    \end{aligned}
\end{equation}
and with an Hopf algebra structure given by:
\begin{equation}\label{hopfbicross}
\begin{aligned}
\mu ((x\otimes a),(y\otimes b)) &=xy_{(1)}\otimes (a\triangleleft y_{(2)})b,\\
1_{\mathcal{X} \vartriangleright \! \blacktriangleleft \mathcal{A}} &=1_\mathcal{X}\otimes 1_\mathcal{A},\\
\Delta(x\otimes a) &=(x_{(1)}\otimes {x_{(2)}}^{(\bar{1})}a_{(1)})\otimes ({x_{(2)}}^{(\bar{2})}\otimes a_{(2)}),\\
\varepsilon (x\otimes a) &=\varepsilon (x)\varepsilon (a),\\
S(x\otimes a) &=(1_{\mathcal{X}} \otimes S(x^{(\bar{1})}a)) \cdot (S(x^{(\bar{2})})\otimes 1_{\mathcal{A}}),
\end{aligned}
\end{equation}
where $x,y \in \mathcal{X}$, $a,b\in \mathcal{A}$ and we employed the Sweedler notation $\Delta (h) = h_{(1)} \otimes h_{(2)}$ and $\beta (x)=x^{(\bar{1})}\otimes x^{(\bar{2})}$, with $x^{(\bar{1})} \in \mathcal{A}$ and $x^{(\bar{2})} \in \mathcal{X}$.

The structure maps must also satisfy the following compatibility conditions with the Hopf structure of $\mathcal{X} \otimes \mathcal{A}$
\begin{equation}\label{comp}
\begin{aligned}
\varepsilon (a\triangleleft x) &=\varepsilon (a)\varepsilon (x),\\
\Delta (a\triangleleft x) &=(a_{(1)}\triangleleft x_{(1)}){x_{(2)}}^{(\bar{1})}\otimes (a_{(2)}\triangleleft {x_{(2)}}^{(\bar{2})}),\\
\beta (xy) &=(x^{(\bar{1})}\triangleleft y_{(1)}){y_{(2)}}^{(\bar{1})}\otimes x^{(\bar{2})}{y_{(2)}}^{(\bar{2})},\\
{x_{(1)}}^{(\bar{1})}(a\triangleleft x_{(2)})\otimes {x_{(1)}}^{(\bar{2})} &=(a\triangleleft x_{(1)}){x_{(2)}}^{(\bar{1})}\otimes {x_{(2)}}^{(\bar{2})}.
\end{aligned}
\end{equation}

We have denoted this structure by means of the symbol $ \vartriangleright \! \blacktriangleleft$; this has been done in analogy with the standard semidirect product, since in our case we have the Hopf algebra $\mathcal{X}$ acting on the Hopf algebra $\mathcal{A}$ and a new kind of dual action, a left coaction, of the algebra $\mathcal{A}$ on $\mathcal{X}$.

From the first equation in~\eqref{hopfbicross} we can see that the bicrossproduct algebra $\mathcal{X} \vartriangleright \! \blacktriangleleft \mathcal{A}$ can always be seen as the universal enveloping algebra generated by elements $X=x\otimes 1$, $\mathcal{A}=1\otimes a$, modulo the commutation relations
\begin{equation}
[X,A]=x\otimes a -x_{(1)}\otimes (a\triangleleft x_{(2)}). \label{1.7}
\end{equation}

\subsection{Bicrossproduct Poincar\'e QGs}\label{bicrossqg}
The main claim of the present Section is that $\mathcal{C}_\varrho (P)$ defined in \eqref{Crho} has a bicrossproduct structure
\begin{equation}\label{bicrosstructure}
\mathcal{C}_\varrho(P) =\mathcal{T}_\varrho^* \vartriangleright \! \blacktriangleleft \mathcal{C}(SO(1,3)),
\end{equation}
where $\mathcal{C}(SO(1,3))$ is the undeformed (Hopf) algebra of continuous functions over the Lorentz group 
and $\mathcal{T}_\varrho^*$ is given by\footnote{Note, here and in the following, the abuse of notation in calling $a^\mu$ objects in $\mathcal{T}_\varrho^*$ that are lifted to $a^\mu\otimes 1\equiv a^\mu$ in $\mathcal{C}_\varrho(P)$.}:
\begin{equation}
\begin{aligned}
\relax[a^\mu, a^\nu] &=i \varrho [{\delta^\nu}_0 (a^2 {\delta^\mu}_1-a^1 {\delta^\mu}_2) -{\delta^\mu}_0 (a^2 {\delta^\nu}_1-a^1 {\delta^\nu}_2)],\\
\Delta (a^\mu) &=a^\mu \otimes 1+1\otimes a^\mu,\\
S(a^\mu) &=-a^\mu,\\
\varepsilon(a^\mu) &=0.
\end{aligned}
\end{equation}
This last algebra is isomorphic to the trivial Hopf algebra extension of~\eqref{algebra}, $\mathcal{M}^H_\varrho$, so that, if our statement is correct, we clearly understand in which sense we are generalising the semidirect product structure to the Hopf algebraic case:
\begin{equation}
    \mathcal{M}^H_\varrho\cong \mathcal{C}_\varrho(P)/_\varrho \mathcal{C}(SO(1,3)),
\end{equation}
where the symbol $/_\varrho$ means that we are taking the quotient with respect to this new action/coaction structure. In the recent paper~\cite{Fabiano:2023uhg} that inspired this work, we gave an explicit proof for this statement. 

We will present, now, a generalisation of the construction that works for general Lie algebra-type spacetimes\footnote{We became aware, after the submission of the first draft of this article, of another paper independently stating this same result \cite{Mercati:2024rzg}.}.
Consider a spacetime with a Lie algebra-type noncommutativity described by:
\begin{equation}
\Theta^{\mu \nu} = {{\Theta^{\mu \nu}}_ \rho}^{(1)} x^\rho \equiv {\theta^{\mu \nu}}_ \rho x^\rho,
\end{equation}
namely,
\begin{equation}
    [x^\mu,x^\nu]=i\alpha {\theta^{\mu\nu}}_\rho x^\rho,
\end{equation}
such that the corresponding classical $r$-matrices can be given by
\begin{equation}
    r=i\frac{\alpha}{2} \theta^{\mu\nu\rho} P_\mu \wedge M_{\nu\rho}.\label{liealgebrar}
\end{equation}

From this, we can write the full QG structure of the associated isometry groups $\mathcal{C}_\alpha(P)$ as
\begin{equation}\label{BB.8}
\begin{aligned}
[a^\mu,a^\nu]&=i\frac{\alpha}{2} \theta^{\nu\gamma\delta} ({\delta^\mu}_\gamma a_\delta -{\delta^\mu}_\delta a_\gamma) +i\frac{\alpha}{2} \theta^{\mu\gamma\delta} ({\delta^\nu}_\delta a_\gamma -{\delta^\nu}_\gamma a_\delta),\\
[{\Lambda^\mu}_\sigma,{\Lambda^\nu}_\rho]&=0,\\
[a^\mu,{\Lambda^\nu}_\rho]&=i \frac{\alpha}{2} \theta^{\lambda\gamma\delta} {\Lambda^\mu}_\lambda (g_{\delta \rho} {\Lambda^\nu}_\gamma - g_{\gamma \rho}{\Lambda^\nu}_\delta )+i\frac{\alpha}{2} \theta^{\mu\gamma\delta}  ({\delta^\nu}_\delta \Lambda_{\gamma \rho} -{\delta^\nu}_\gamma \Lambda_{\delta \rho}),\\
\Delta(a^\mu)&={\Lambda^\mu}_\alpha \otimes a^\alpha +a^\mu \otimes 1,\\
\Delta({\Lambda^\mu}_\nu)&={\Lambda^\mu}_\alpha \otimes {\Lambda^\alpha}_\nu,\\
\varepsilon(a^\mu)&=0,\\
\varepsilon({\Lambda^\mu}_\nu)&={\delta^\mu}_\nu,\\
S(a^\mu)&=-a^\nu {(\Lambda^{-1})^\mu}_\nu,\\
S({\Lambda^\mu}_\nu)&={(\Lambda^{-1})^\mu}_\nu.
\end{aligned}
\end{equation}

Our claim is, then, that these Lie algebra-type quantum groups $\mathcal{C}_\alpha(P)$ have a canonical explicit bicrossproduct structure given by
\begin{equation}\label{bicrossgroup}
\mathcal{C}_\alpha(P) =\mathcal{T}_\alpha^* \vartriangleright \! \blacktriangleleft \mathcal{C}(SO(1,3)),
\end{equation}
with $\mathcal{C}(SO(1,3))$ being the Hopf algebra of continuous function on the classical Lorentz group, and $\mathcal{T}_\alpha^*$ the Hopf algebra of translations defined by
\begin{equation}
\begin{aligned}
[a^\mu, a^\nu] &=i\frac{\alpha}{2} \theta^{\nu\gamma\delta} ({\delta^\mu}_\gamma a_\delta -{\delta^\mu}_\delta a_\gamma) +i\frac{\alpha}{2} \theta^{\mu\gamma\delta} ({\delta^\nu}_\delta a_\gamma -{\delta^\nu}_\gamma a_\delta),\\
\Delta (a^\mu) &=a^\mu \otimes 1+1\otimes a^\mu,\\
S(a^\mu) &=-a^\mu,\\
\varepsilon(a^\mu) &=0.
\end{aligned}
\end{equation}
The action and coaction maps are given as
\begin{equation}
\begin{aligned}
    {\Lambda^\nu}_\rho \triangleleft a^\mu&=[{\Lambda^\nu}_\rho, a^\mu],\\
    \beta(a^\mu)&={\Lambda^\mu}_\nu \otimes a^\nu.
\end{aligned}
\end{equation}

Given this construction, it is possible to show that all the properties of a bicrossproduct Hopf algebra are satisfied, following the same kind of proof reported in~\cite{Fabiano:2023uhg}.
For the $\kappa$-case, a bicrossproduct structure in the form \eqref{bicrossgroup} with a structure given by the previous construction was found in \cite{Majid:1994cy}, while we found it for the $\varrho$-case in \cite{Fabiano:2023uhg}.

We stress that this construction works for Lie algebra-type spacetimes and not, for instance, for canonical ones.

\section{The \texorpdfstring{$\varrho$}{}-Poincar\'e QUEA}\label{sec:QUEA}
It is well known that dual to a quantum group obtained deforming the algebra of continuous functions over a Lie group $G$, there exists another quantum group, often called a quantum universal enveloping algebra (QUEA), obtained deforming the Hopf structure of the universal enveloping algebra of the corresponding Lie algebra $\mathfrak{g}$. We can, in general, choose different sets of generators in order to express the structure of a QUEA and these ``bases'' are connected one to the other via nonlinear transformations of the generators. The most famous example in which this feature comes into play is the $\kappa$-Poincar\'e case, whose QUEA is often studied in two notable bases: the ``standard'' and the ``Majid-Ruegg'' (or ``bicrossproduct'') one \cite{1994qkp..rept.....L}.
In this section we will start introducing the $\varrho$-Poincar\'e QUEA, $U_\varrho(\mathfrak{p})$, in the most simple way possible, employing a well-established procedure called ``twisting'' that will lead us to the ``twisted'' QUEA basis. Once this is done, we will turn to another basis in which the bicrossproduct nature of this QUEA is fully manifest.

\subsection{\texorpdfstring{$U_\varrho(\mathfrak{p})$}{} in the ``twisted'' basis}
If we have a Lie algebra, $\mathfrak{g}$, with an associated classical $r$-matrix whose carrier algebra is abelian, then the model obey to a CYBE and we call its deformation done through $r$ a ``soft'' or ``abelian'' deformation.
In this case there exists a simple way to quantise the Hopf structure of its universal enveloping algebra, $U(\mathfrak{g})$. This is done deforming the coproducts and antipodes through a map
\begin{equation}
    \mathcal{F}\in U(\mathfrak{g})\otimes U(\mathfrak{g}),
\end{equation}
called a Drinfel'd twist of the \textit{Reshetikhin} type (for further details on the procedure see, for instance, \cite{Lukierski:1993hk}).
            
The Drinfel'd twist for the $\varrho$-Minkowski case was introduced in~\cite{Lukierski}, and reads
\begin{equation}\label{rhotwist}
\mathcal{F}=e^{\frac{i\varrho}{2}[P_0\wedge M_{12}]}.
\end{equation}
It is trivial to check how the first order in the $\varrho$-expansion of this object reproduces the classical $r$-matrix~\eqref{rrho}.

From now on, we will work with the standard redefinition of the Lorentz generators $R_i=\frac{1}{2}\epsilon_{ijk}M_{jk}$, $N_i=M_{i0}$.
The QUEA corresponding to this twist, $U_\varrho(\mathfrak{p})$, was first introduced in~\cite{Ciric:2017rnf} and it is given deforming the coproducts of the undeformed Poincar\'e Hopf algebra to
\be\label{twistedrhoqg}
\begin{array}{r@{}l}
    \Delta_\mathcal{F}P_0 =&P_0\otimes 1+1\otimes P_0,\\
\Delta_\mathcal{F}P_1 =&P_1\otimes \cos \left(\frac{\varrho}{2}P_0 \right)+\cos \left(\frac{\varrho}{2}P_0 \right)\otimes P_1+P_2\otimes \sin \left(\frac{\varrho}{2}P_0 \right)-\sin \left(\frac{\varrho}{2}P_0 \right)\otimes P_2,\\
\Delta_\mathcal{F}P_2 =&P_2\otimes \cos \left(\frac{\varrho}{2}P_0 \right)+\cos \left(\frac{\varrho}{2}P_0 \right)\otimes P_2-P_1\otimes \sin \left(\frac{\varrho}{2}P_0 \right)+\sin \left(\frac{\varrho}{2}P_0 \right)\otimes P_1,\\
\Delta_\mathcal{F}P_3 =&P_3\otimes 1+1\otimes P_3,\\
\Delta_\mathcal{F}R_{1} =&R_{1}\otimes\cos \left(\frac{\varrho}{2}P_0 \right)+\cos \left(\frac{\varrho}{2}P_0 \right)\otimes R_{1}+R_{2}\otimes\sin \left(\frac{\varrho}{2}P_0 \right)-\sin \left(\frac{\varrho}{2}P_0 \right)\otimes R_{2},\\
\Delta_\mathcal{F}R_{2} =&R_{2}\otimes\cos \left(\frac{\varrho}{2}P_0 \right)+\cos \left(\frac{\varrho}{2}P_0 \right)\otimes R_{2}-R_{1}\otimes\sin \left(\frac{\varrho}{2}P_0 \right)+\sin \left(\frac{\varrho}{2}P_0 \right)\otimes R_{1},\\
\Delta_\mathcal{F}R_{3} =&R_{3}\otimes 1+1\otimes R_{3},\\
\Delta_\mathcal{F}N_{1} =&N_{1}\otimes\cos \left(\frac{\varrho}{2}P_0 \right)+\cos \left(\frac{\varrho}{2}P_0 \right)\otimes N_{1}+N_{2}\otimes\sin \left(\frac{\varrho}{2}P_0 \right)-\sin \left(\frac{\varrho}{2}P_0 \right)\otimes N_{2}\\
&+P_1\otimes \frac{\varrho}{2}R_{3}\cos \left(\frac{\varrho}{2}P_0 \right)-\frac{\varrho}{2}R_{3}\cos \left(\frac{\varrho}{2}P_0 \right)\otimes P_1+P_2\otimes \frac{\varrho}{2}R_{3}\sin \left(\frac{\varrho}{2}P_0 \right)\\
&+\frac{\varrho}{2}R_{3}\sin \left(\frac{\varrho}{2}P_0 \right)\otimes P_2,\\
\Delta_\mathcal{F}N_{2} =&N_{2}\otimes\cos \left(\frac{\varrho}{2}P_0 \right)+\cos \left(\frac{\varrho}{2}P_0 \right)\otimes N_{2}-N_{1}\otimes\sin \left(\frac{\varrho}{2}P_0 \right)+\sin \left(\frac{\varrho}{2}P_0 \right)\otimes N_{1}\\
&+P_2\otimes \frac{\varrho}{2}R_{3}\cos \left(\frac{\varrho}{2}P_0 \right)-\frac{\varrho}{2}R_{3}\cos \left(\frac{\varrho}{2}P_0 \right)\otimes P_2-P_1\otimes \frac{\varrho}{2}R_{3}\sin \left(\frac{\varrho}{2}P_0 \right)\\
&-\frac{\varrho}{2}R_{3}\sin \left(\frac{\varrho}{2}P_0 \right)\otimes P_1,\\
\Delta_\mathcal{F}N_{3} =&N_{3}\otimes 1+1\otimes N_{3}+\frac{\varrho}{2}P_3\otimes R_{3}-\frac{\varrho}{2}R_{3}\otimes P_3.
\end{array}
\ee

We would like, at this point, to define a bicrossproduct splitting of this QUEA that ``disentangles'' the translational sector from the Lorentz one in the following way:
\begin{equation}
    U_\varrho(\mathfrak{p})=U(\mathfrak{so}(1,3)) \vartriangleright \! \blacktriangleleft T_\varrho,
\end{equation}
where $U(\mathfrak{so}(1,3))$ is the undeformed universal enveloping algebra of the Lorentz algebra and $T_\varrho$ a translation Hopf algebra with deformed coproducts isomorphic to the first four maps in~\eqref{twistedrhoqg}. Alas, as it can be checked explicitly, it is not possible in this way to define a good left coaction compatible with all the requirements of Section~\ref{sec:bicrossHopf}. On the other hand, by Hopf duality, we know that if the $\varrho$-Poincar\'e QG has a bicrossproduct structure, then also the QUEA has to have it. We will see, in the next section, how to solve this apparent issue.
 
\subsection{\texorpdfstring{$U_\varrho(\mathfrak{p})$}{} in the ``bicrossproduct'' basis}\label{sec:bicrossU}
By analogy with the known procedure for the $\kappa$-case~\cite{Majid:1994cy}, we perform the following non-linear transformation of the generators:
\be\label{redef}
\begin{array}{r@{}l}
    \widetilde{P}_0&=P_0,\\
    \widetilde{P}_1&=P_1\cos(\frac{\varrho}{2}P_0)-P_2\sin(\frac{\varrho}{2}P_0), \\
    \widetilde{P}_2&=P_2\cos(\frac{\varrho}{2}P_0)+P_1\sin(\frac{\varrho}{2}P_0), \\
    \widetilde{P}_3&=P_3,\\
    \widetilde{R}_1&=R_1\cos(\frac{\varrho}{2}P_0)-R_2\sin(\frac{\varrho}{2}P_0), \\
    \widetilde{R}_2&=R_2\cos(\frac{\varrho}{2}P_0)+R_1\sin(\frac{\varrho}{2}P_0), \\
    \widetilde{R}_3&=R_3,\\
    \widetilde{N}_1&={N}_1\cos(\frac{\varrho}{2}P_0)-{N}_2\sin(\frac{\varrho}{2}P_0)+\frac{\varrho}{2}R_3 \widetilde{P}_1, \\
    \widetilde{N}_2&={N}_2\cos(\frac{\varrho}{2}P_0)+{N}_1\sin(\frac{\varrho}{2}P_0)+\frac{\varrho}{2}R_3 \widetilde{P}_{2},\\
    \widetilde{N}_3&={N}_3+\frac{\varrho}{2}R_3P_3.
\end{array}
\ee

By direct calculations it is possible to check that the algebra sector is still undeformed in these new generators, while for the coproducts we have
\be\label{bicrosscoprod}
\begin{array}{r@{}l}
    &\Delta \widetilde{P}_0=\widetilde{P}_0\otimes 1+1\otimes \widetilde{P}_0,\\
    &\Delta \widetilde{P}_1=\widetilde{P}_1\otimes 1 +\cos(\varrho \widetilde{P}_0)\otimes \widetilde{P}_1-\sin(\varrho \widetilde{P}_0)\otimes \widetilde{P}_2,\\
    &\Delta \widetilde{P}_2=\widetilde{P}_2\otimes 1 +\cos(\varrho \widetilde{P}_0)\otimes \widetilde{P}_2+\sin(\varrho \widetilde{P}_0)\otimes \widetilde{P}_1,\\
    &\Delta \widetilde{P}_3=\widetilde{P}_3\otimes 1+1\otimes \widetilde{P}_3,\\
    &\Delta \widetilde{R}_1=\widetilde{R}_1\otimes 1 +\cos(\varrho \widetilde{P}_0)\otimes \widetilde{R}_1-\sin(\varrho \widetilde{P}_0)\otimes \widetilde{R}_2,\\
    &\Delta \widetilde{R}_2=\widetilde{R}_2\otimes 1 +\cos(\varrho \widetilde{P}_0)\otimes \widetilde{R}_2+\sin(\varrho \widetilde{P}_0)\otimes \widetilde{R}_1,\\ 
    &\Delta \widetilde{R}_3=\widetilde{R}_3\otimes 1+1\otimes \widetilde{R}_3,\\
    &\Delta \widetilde{N}_1=\widetilde{N}_1\otimes 1 +\cos(\varrho \widetilde{P}_0)\otimes \widetilde{N}_1-\sin(\varrho \widetilde{P}_0)\otimes \widetilde{N}_2+\varrho \widetilde{P}_1\otimes \widetilde{R}_3,\\
    &\Delta \widetilde{N}_2=\widetilde{N}_2\otimes 1 +\cos(\varrho \widetilde{P}_0)\otimes \widetilde{N}_2+\sin(\varrho \widetilde{P}_0)\otimes \widetilde{N}_1+\varrho \widetilde{P}_2\otimes \widetilde{R}_3,\\
    &\Delta \widetilde{N}_3=\widetilde{N}_3\otimes 1 +1\otimes \widetilde{N}_3+\varrho \widetilde{P}_3\otimes \widetilde{R}_3.
\end{array}
\ee

The counits are still trivial
\begin{equation}\varepsilon(\widetilde{P}_\mu)=\varepsilon(\widetilde{N}_i)=\varepsilon(\widetilde{R}_i)=0,
\end{equation}
an the antipodes are changed accordingly to
\be\begin{array}{r@{}l}\label{bicrossantip}
    &S(\widetilde{P}_0)=-\widetilde{P}_0, \\
    &S(\widetilde{P}_1)=-\widetilde{P}_1\cos(\varrho \widetilde{P}_0)-\widetilde{P}_2\sin(\varrho \widetilde{P}_0), \\ &S(\widetilde{P}_2)=-\widetilde{P}_2\cos(\varrho \widetilde{P}_0)+\widetilde{P}_1\sin(\varrho \widetilde{P}_0), \\
    &S(\widetilde{P}_3)=-\widetilde{P}_3, \\
    &S(\widetilde{R}_1)=-\widetilde{R}_1\cos(\varrho \widetilde{P}_0)-\widetilde{R}_2\sin(\varrho \widetilde{P}_0), \\ &S(\widetilde{R}_2)=-\widetilde{R}_2\cos(\varrho \widetilde{P}_0)+\widetilde{R}_1\sin(\varrho \widetilde{P}_0),\\
    &S(\widetilde{R}_3)=-\widetilde{R}_3, \\
    &S(\widetilde{N}_1)=-\cos(\varrho \widetilde{P}_0)\widetilde{N}_1-\sin(\varrho \widetilde{P}_0)\widetilde{N}_2+\varrho\cos(\varrho \widetilde{P}_0)\widetilde{P}_1\widetilde{R}_3+\varrho\sin(\varrho \widetilde{P}_0)\widetilde{P}_2\widetilde{R}_3,\\
    &S(\widetilde{N}_2)=-\cos(\varrho \widetilde{P}_0)\widetilde{N}_2+\sin(\varrho \widetilde{P}_0)\widetilde{N}_1+\varrho\cos(\varrho \widetilde{P}_0)\widetilde{P}_2\widetilde{R}_3-\varrho\sin(\varrho \widetilde{P}_0)\widetilde{P}_1\widetilde{R}_3,\\
    &S(\widetilde{N}_3)=-\widetilde{N}_3+\varrho \widetilde{R}_3\widetilde{P}_3.
\end{array}
\ee

The quantum Hopf algebra    $U_\varrho(\mathfrak{p})$ admits, then, the bicrossproduct decomposition 
\begin{equation}
U_\varrho(\mathfrak{p}) =U(\mathfrak{so}(1,3))\vartriangleright \! \blacktriangleleft \mathcal{T}_\varrho,
\end{equation}
with $\mathcal{T}_\varrho$ the new deformed translation sector obtained keeping trivial commutation relations and counits and modifying coproducts and antipodes as in~\eqref{bicrosscoprod} and~\eqref{bicrossantip}. The full details of the proof can be found in~\cite{Fabiano:2023uhg}, and will be omitted here for brevity.

Let us close this section noting that all the QUEA dual to the Lie algebra-type QGs discussed in Section~\ref{bicrossqg} have (by duality) a bicrossproduct structure, even though it is in general not trivial to find a suitable change of generators in which this feature is manifest.

\subsection{Bicrossproduct as a generalisation of the semidirect product}
We mentioned before, in Section~\ref{sec:bicrossHopf}, that the bicrossproduct structure is a generalisation of the notion of a semidirect product in the context of Hopf algebras. In this subsection we will elaborate on this matter.

At the classical level, a semidirect sum $\mathfrak{g}=\mathfrak{h} \oplus_s \mathfrak{f}$ of lie algebras $\mathfrak{h}, \mathfrak{f}$ is a split extension  
\begin{equation}
    \mathfrak{h} \xhookrightarrow{i} \mathfrak{g} \xrightarrow{\pi} \mathfrak{f},
\end{equation}
$i$ being a monomorphism and a $\pi$ an epimorphism. The same holds for Lie groups. 
A Lie group $G$  is  a semidirect product of $H$ and $F$, namely $G=H\rtimes F$, if
\begin{equation}
    H\xhookrightarrow{} G\xrightarrow{} F,
\end{equation}
is a split short exact sequence.

In~\cite{Majid1990PhysicsFA,matched,singer} the bicrossproduct structure was introduced as a generalisation of this construction to the Hopf algebraic context. Furthermore, in~\cite{Majid:1994cy}, this procedure was applied to the  $\kappa$-Poincar\'e quantum group $U_\kappa(\mathfrak{p})$ in the ``Majid-Ruegg basis'' and the Hopf algebra-isomorphism with the ``standard basis'' description was explicitly described. 
In~\cite{Fabiano:2023uhg} the procedure was applied to the $\varrho$-Poincar\'e case. We will report, here, the basic steps.

We can see the bicrossproduct as resulting from the following short exact sequence structure:
\be
\begin{aligned}
\mathcal{T}_\varrho\xhookrightarrow{i}U_\varrho(\mathfrak{p})\xrightarrow{\pi} U(\mathfrak{so}(1,3)),\\
\mathcal{T}_\varrho\xleftarrow{p}U_\varrho(\mathfrak{p})\xhookleftarrow{j} U(\mathfrak{so}(1,3)),
\end{aligned}
\ee
where $i$ is a monomorphism, $\pi$ an epimorphism, $j$ is an algebra homomorphism and $p$ is a coalgebra homomorphism, such that:
\begin{equation}\label{strucmaps}
\begin{aligned}
    &\pi(\widetilde{N}_i)=n_i, && \pi(\widetilde{R}_i)=m_i, \hspace{0.5cm}&& \pi(\widetilde{P}_\mu)=0,\\
    &\pi\circ j= id, && p\circ i=id,&&\\
    &(p\otimes p)\circ \Delta= \Delta\circ p, \hspace{0.5cm}&& \varepsilon\circ p= p,&&\\
    &(id\otimes j)\circ \Delta = (\pi\otimes id)\circ \Delta \circ j,\hspace{0.5cm}&&p(u)t=p(u\, i(t)),&&
\end{aligned}
\end{equation}
where $u\in U_\varrho(\mathfrak{p}), t\in\mathcal{T}_\varrho$ and $n_i, m_i\in U(\mathfrak{so}(1,3))$ are the classical boost and rotation generators.

Taking
\begin{align}\label{appareepsilon}
    \widetilde{P}_0&\coloneqq i(P_0)=P_0,\nonumber\\
    \widetilde{P}_A&\coloneqq i(P_A)=P_A\cos(\frac{\varrho}{2}P_0)-\epsilon_{AB}P_B\sin(\frac{\varrho}{2}P_0),\nonumber\\
    \widetilde{P}_3&\coloneqq i(P_3)=P_3,\nonumber\\
    \widetilde {R}_A &\coloneqq  j(m_A)=R_A\cos(\frac{\varrho}{2}P_0)-\epsilon_{AB}R_B\sin(\frac{\varrho}{2}P_0), \\
    \widetilde R_3 &\coloneqq  j(m_3)=R_3,\nonumber\\
    \widetilde N_A &\coloneqq  j(n_A)=N_A\cos(\frac{\varrho}{2}P_0)-\epsilon_{AB}N_B\sin(\frac{\varrho}{2}P_0)
    +\frac{\varrho}{2}R_3\left(P_A\cos(\frac{\varrho}{2}P_0)-\epsilon_{AB}P_B\sin(\frac{\varrho}{2}P_0)\right),\nonumber\\
    \widetilde N_3 &\coloneqq  j(n_3)=N_3+\frac{\varrho}{2}R_3P_3,\nonumber
\end{align}
where $\epsilon_{AB}\,, A,B=1,2$ is the Levi-Civita pseudotensor in $2$ dimensions, one can verify that the relevant properties of~\eqref{strucmaps} are satisfied. Comparing with~\eqref{redef}, we see that the maps $i, j$ give the nonlinear change of Lorentz generators that we have performed to go from the ``twisted'' basis to the ``bicrossproduct'' one. 

The action and coaction maps can be defined as  
\be
\begin{aligned}
    t\triangleleft h &= j(Sh_{(1)})tj(h_{(2)}), \\
    \beta(\pi(u))&= p(u_{(1)})Sp(u_{(3)})\otimes\pi(u_{(2)}),
\end{aligned}
\ee
with $h\in U(\mathfrak{so}(1,3))$, and $\Delta^2a=a_{(1)}\otimes a_{(2)} \otimes a_{(3)}$. In this way we obtained a bicrossproduct Hopf algebra whose commutators and coproducts are constructed as
\begin{equation} \label{hopstruc}
    \begin{aligned}
    i(t)j(h)&=j(h_{(1)})i(t\triangleleft h_{(2)}),\\
    \Delta(i(t))&=i(t_{(1)})\otimes i(t_{(2)}),\\ 
    \Delta (j(h))&=j(h_{(1)})(i\otimes j)\circ \beta(h_{(2)}),
    \end{aligned}
\end{equation}
reproducing the last subsection structure maps.

\section{\texorpdfstring{$\star$}{}-products in \texorpdfstring{$\varrho$}{}-Minkowski}\label{sec:star}
Starting from the commutation rules of spacetime one can represent the algebra of noncommutative functions as an algebra of operators by choosing a specific ordering for   the basis of noncommutative plane waves $\hat\phi(p)=e^{i p_\mu \hat x^\mu}$~\cite{Amelino-Camelia:1999jfz}.   
The operator product  of   plane waves $\hat \phi(p)$  defines a noncommutative $\star$-product for functions on spacetime according to
\be \label{starp}
(f\star g)(x) =\int d^4 p \, d^4 k\, \Tilde{f}(p)\Tilde{g}(k) \; \rm{e}^{ip_\mu x^\mu}
\star  \rm{e}^{ik_\nu x^\nu},
\ee
where the tilde indicates the standard Fourier transform. Different ordering prescriptions will define different products.
The $\star$-product between exponentials can be rewritten as
\be\label{expprod}
e^{ip\cdot x}\star e^{ik\cdot x}=e^{i(p\oplus k)\cdot x},
\ee
$\oplus$ representing a deformed sum. It is well known that this defines a  coproduct for the translation generators (see for instance~\cite{agostini} where it was applied to the case of $\kappa$-Poincar\'e), derived imposing consistency of the action of translations  on both sides of \eqn{expprod}.

In order to define plane waves and their rule of multiplication, we start with  the following realization of the spacetime algebra of the commutation relations~\eqref{algebra}~\cite{Freidel:2007hk, Kowalski-Glikman:2013rxa}:
\begin{equation}\small
\label{corrrep}
    \hat x^0=\begin{pmatrix}
        0 & i\varrho & 0 & 0 \\
        -i\varrho & 0 & 0 & 0 \\
        0 & 0 & 0 & 0 \\
        0 & 0 & 0 & 0
    \end{pmatrix} , \quad 
      \hat x^1=\begin{pmatrix}
        0 & 0 & 0 & i\varrho  \\
        0 & 0 & 0 & 0 \\
        0 & 0 & 0 & 0 \\
        0 & 0 & 0 & 0
    \end{pmatrix} , \quad
        \hat x^2=\begin{pmatrix}
        0 & 0 & 0 & 0  \\
        0 & 0 & 0 & i\varrho \\
        0 & 0 & 0 & 0 \\
        0 & 0 & 0 & 0
    \end{pmatrix} ,\quad 
            \hat x^3=\begin{pmatrix}
        0 & 0 & 0 & 0  \\
        0 & 0 & 0 & 0 \\
        0 & 0 & 0 & i\varrho \\
        0 & 0 & 0 & 0
    \end{pmatrix}.
\end{equation}

\subsection{Time-to-the-right ordering and bicrossproduct basis}
We start analysing the time-to-the-right ordering. By means of~\eqref{corrrep}, we can write
\begin{equation}
\label{pwr}
    \hat \phi_R(p)=e^{ip_k\hat x^k}e^{ip_0\hat x^0}= \begin{pmatrix}
    \cos(\varrho p_0) & -\sin(\varrho p_0) & 0 & -\varrho p_1 \\
    \sin(\varrho p_0) & \cos(\varrho p_0) & 0 & -\varrho p_2 \\
    0 & 0 & 1 & -\varrho p_3 \\
    0 & 0 & 0 & 1
    \end{pmatrix}.
\end{equation}
Multiplying two plane waves we get
\begin{equation}
\begin{aligned}
\label{ttrsum}
    &\hat\phi_R(p)\hat\phi_R(k)=\hat \phi_R(p\oplus_R k)=\\
   & =\begin{pmatrix}
    \cos(\varrho(k_0+p_0)) & -\sin(\varrho(k_0+p_0)) & 0 & -\varrho [p_1+k_1\cos(\varrho p_0)-k_2\sin(\varrho p_0)] \\
    \sin(\varrho(k_0+p_0)) & \cos(\varrho(k_0+p_0)) & 0 & -\varrho[p_2+k_2\cos(\varrho p_0)+k_1\sin(\varrho p_0)] \\
    0 & 0 & 1 & -\varrho(p_3+k_3) \\
    0 & 0 & 0 & 1 
    \end{pmatrix}.
\end{aligned}
\end{equation}
Comparing with~\eqref{pwr}, we define the deformed sum as
\begin{equation}
\label{complaws}
    \begin{cases}
    (p\oplus_R k)_0= p_0+k_0,\\
    (p\oplus_R k)_1= p_1+k_1\cos(\varrho p_0)-k_2\sin(\varrho p_0),\\
    (p\oplus_R k)_2= p_2 + k_2\cos(\varrho p_0)+k_1\sin(\varrho p_0), \\
    (p\oplus_R k)_3=p_3+k_3.
    \end{cases}
\end{equation}

The translation generators act as differential operators on exponentials: $P_\mu^R \phi(p)=-i\partial_\mu e^{i p_\nu x^\nu}=p_\mu \phi(p)$. Therefore we have 
\be
P^R_\mu \left(\phi(p)\star\phi(k)\right)= -i \partial_\mu \exp (i(p\oplus_R k)_\nu x^\nu)= (p\oplus_R k)_\mu \phi(p\oplus_R k)=\mu \circ \Delta P^R_\mu (\phi(p)\otimes \phi(k)).
\ee
Consistency with the above requirement implies
 \be\label{coproductspr}
 \begin{array}{r@{}l}
    &\Delta P_0^R=P_0^R\otimes 1+1\otimes P_0^R,\\
    &\Delta P_1^R=P_1^R\otimes 1 +\cos(\varrho P_0^R)\otimes P_1^R-\sin(\varrho P_0^R)\otimes P_2^R,\\
    &\Delta P_2^R=P_2^R\otimes 1 +\cos(\varrho P_0^R)\otimes P_2^R+\sin(\varrho P_0^R)\otimes P_1^R,\\
    &\Delta P_3^R=P_3^R\otimes 1+1\otimes P_3^R\,,
    \end{array}
    \ee
that are the coproducts we found in the bicrossproduct basis.

We are, then, ready to look at relevant features of the $\star$-product associated with the time-to-the-right ordering.
We find 
\bea
( f\star g) (x) = \int d^4 p\;  (\Tilde f\circ \Tilde g)(p)e^{i p_\mu x^\mu}, \label{starpr}
\eea
with  
\be 
(\Tilde f\circ \Tilde g)(p) =\int d^4  k \; \Tilde f(p_0-k_0,p_A-R_{AB}(-\varrho p_0+\varrho k_0)k_B,p_3-k_3)  \;\Tilde g(k),
\ee
the deformed convolution of Fourier transforms, where $R_{AB}(\theta), \; A,B=1,2$, is the rotation matrix in the 1-2 plane of argument $\theta$.

By direct calculation it is easy to check that this product is cyclic with respect to the standard integration measure on $\mathbb{R}^4$:
\begin{equation}
    \int d^4{x}\; f({x})\star g({x}) =\int d^4{x}\; g({x})\star f({x}),
\end{equation}
and this allows to build $\star$-gauge invariant actions in field theories on $\mathcal{M}_\varrho$.

However, the closure condition \cite{Felder:2000nc} 
\begin{equation}
    \int d^4{x}\; f({x})\star g({x})= \int d^4{x}\; f({x}) \cdot g({x})
\end{equation}
does not hold. This means that quadratic terms in the action, such as the kinetic and the mass term, are, in general, deformed giving modified tree level propagators.

\subsection{Time-symmetric ordering and twist basis}
We turn, now, to the time-symmetric ordering case 
\begin{equation}
\label{pws}
\begin{aligned}\small
    \hat\phi_S(p)=e^{i\frac{p_0\hat x^0}{2}}e^{ip_k\hat x^k}e^{\frac{p_0\hat x^0}{2}}=\begin{pmatrix}
        \cos(\varrho p_0) & -\sin(\varrho p_0) & 0 & -\varrho (p_1\cos(\varrho\frac{p_0}{2})-p_2\sin(\varrho\frac{p_0}{2})) \\
    \sin(\varrho p_0) & \cos(\varrho p_0) & 0 & -\varrho (p_2\cos(\varrho\frac{p_0}{2})+p_1\sin(\varrho\frac{p_0}{2})) \\
    0 & 0 & 1 & -\varrho p_3 \\
    0 & 0 & 0 & 1
    \end{pmatrix}.
\end{aligned}
\end{equation}

Computing the product of symmetric ordered plane waves, as done for the previous case, we obtain the deformed composition law $\oplus_S$:
\begin{equation}
    \begin{cases}
        (p\oplus_S k)_0=p_0+k_0,\\
        (p\oplus_S k)_1=p_1\cos(\frac{\varrho}{2}k_0)+\cos(\frac{\varrho}{2}p_0)k_1+p_2\sin(\frac{\varrho}{2}k_0)-\sin(\frac{\varrho}{2}p_0)k_2,\\
        (p\oplus_S k)_2=p_2\cos(\frac{\varrho}{2}k_0)+\cos(\frac{\varrho}{2}p_0)k_2-p_1\sin(\frac{\varrho}{2}k_0)+\sin(\frac{\varrho}{2}p_0)k_1,\\
        (p\oplus_S k)_3=p_3+k_3.
    \end{cases}
\end{equation}
This is the deformed sum obtained in \cite{DimitrijevicCiric:2018blz} by computing the twisted $\star$-product of plane waves.
Introducing  time-symmetric translation generators, we want to find, as in the previous subsection, the coproduct $\Delta P^S$ which is compatible with the request that $\phi(p\oplus_S k)$ be an eigenfunction of translations $P^S_\mu \phi(p\oplus_S k)=(p\oplus_S k)_\mu\phi(p\oplus_S k )$. These coproducts are the twisted ones~\eqref{twistedrhoqg}.
In~\cite{dimi2} it was found that this $\star$-product is both cyclic and close, and the one-loop behaviour of scalar field theories was there investigated.

An important remark to do is that, as expected, momenta in the time-to-the-right basis are related to the ones in the time-symmetric basis via the same nonlinear transformation that brings the twist basis into the bicrossproduct one:
\begin{equation}
\begin{aligned}
\label{cob}
    \Tilde{p}_1&=p_1\cos(\varrho\frac{p_0}{2})-p_2\sin(\varrho\frac{p_0}{2}),\\ \Tilde{p}_2&=p_2\cos(\varrho\frac{p_0}{2})+p_1\sin(\varrho\frac{p_0}{2}),\\ \Tilde{p}_{0,3}&=p_{0,3}.
\end{aligned}
\end{equation}

Before closing this section let us just note that another different $\star$-product for $\mathcal{M}_\varrho$ was found in~\cite{Hersent:2023lqm} via Weyl quantisation.

\section{Physical bases and the \texorpdfstring{$\varrho$}{}-Poincar\'e QLA}\label{sec:qa}
As we have seen in the previous sections, there seems to be somewhat an ambiguity in the choice of the QUEA basis. This is just an apparent issue, since the QUEA structure is uniquely determined, and all the different bases are just different realizations of its Hopf algebraic structure. In this sense we have a degree of freedom in choosing the basis we prefer for our analyses. However, this is not a novel feature of quantum groups. If, for instance, we consider the classical Poincar\'e algebra we can operate a linear change of generators that mixes the Lorentz and translational ones, such that the abstract algebra is still the Poincar\'e one but it cannot be regarded anymore as a semidirect sum of a translational sector and a Lorentz one. At the QUEA level the discussion is more complex, since now non-linear transformations are allowed. Is there, then, a ``right'' basis to work with, that is preferable over all the others?
To address this question we have to keep in mind that this construction is abstract and to match with real possible experiments one has to assign a physical notion to the objects that appear in the theory. For instance, while for $\varrho$-Poincar\'e both the bases we have dealt with lead to an undeformed algebra sector, and therefore to undeformed quadratic Casimir operators, it is well known that for the $\kappa$-Poincar\'e case bases in which the quadratic Casimir is deformed do exist. From this observation numerous proposals to find quantum gravity effects from a modified mass-shell relation have been made. One can, then, argue that until no new deformed mass-shell effects are found experimentally, good candidates to be physical bases are the ones in which the algebra sector of the QUEA is not deformed, and physical observables are the ones related to these bases. In this picture, it seems very natural to identify bicrossproduct bases with physical ones, since these preserve the classical algebra structure of Poincar\'e generators and provide a generalisation of the splitting between translational and Lorentz sectors. However, this is a very naive expectation, based on an aesthetical principle of minimal deformation from the classical-like structures, and it is in no way a decisive argument.

An interesting observation that has to be made here is that, up to now, we considered a deformation of the universal enveloping algebra of Poincar\'e, and not a direct deformation of the Lie algebra. On the contrary with respect to the classical case, where given a universal enveloping algebra the corresponding Lie algebra that generates it can be easily determined, in the quantum context it is not possible to find a classical Lie algebra that generates the QUEA~\cite{2008JPhCS.128a2049B}. In~\cite{klimyk2012quantum,Aschieri:2005zs,Aschieri:2009zzb} a procedure is described to define a quantum Lie algebra (QLA) describing infinitesimal quantum transformations. This can be viewed as the minimal deformation of a classical Lie algebra, thus defining the quantum analog structure to an ordinary Lie algebra. A concrete possibility, then, would be that physical observables are not related directly to QUEA bases, but to QLA ones~\cite{Aschieri:2017ost}. This construction allows, in particular, to define a well-behaved bicovariant differential calculus \cite{Aschieri:1992wg}.

In order to define a QLA as the minimal deformation of a classical Lie algebra, we require (cfr.~\cite{Aschieri:2009zzb}) that its generators $t_i$ generate the QUEA, that the Leibniz rule is minimally modified as
\begin{equation}
    \Delta(t_i)=t_i\otimes 1+ f_i^j \otimes t_j,
\end{equation}
where $f_i^j$ are in the QUEA, and that there is closure under the deformed adjoint action
\begin{equation}
    [t_i,t_j]=t_{i(1)} t_j S(t_{i(2)})=c_{ij}{}^{k}t_k,
\end{equation}
with $c_{ij}{}^{k}$ structure constants.
If we have a QUEA obtained through a Drinfel'd twist ($\mathcal{F}^{-1}=\overline{f}^\alpha\otimes\overline{f}_\alpha$), the general procedure to obtain this QLA is to twist the generators as
\be
\begin{aligned}\label{qlatrasf}
    P_\mu^{\mathcal{F}}&=\overline{f}^\alpha(P_\mu)\overline{f}_\alpha\,,\\
    M_{\mu\nu}^{\mathcal{F}}&=\overline{f}^\alpha(M_{\mu\nu})\overline{f}_\alpha\,.
\end{aligned}
\ee

Applying now this procedure to $U_\varrho(\mathfrak{p})$ in the twist basis, one explicitly finds the nonlinear change of generators \eqref{redef}. This means that the generators in the twist basis, along with the structure maps given in Section~\ref{sec:bicrossU}, define the $\varrho$-Poincar\'e quantum Lie algebra $\mathfrak{iso}_\varrho(1,3)=\mathfrak{p}_\varrho$. It is interesting to note that the nonlinear change of generators we found by comparison with the $\kappa$ case can be formally described by the transformation~\eqref{qlatrasf}, thus apparently connecting the QLA to the QUEA in the bicrossproduct basis.

\section{Conclusions}\label{sec:concl}
In this work we reviewed the construction made in~\cite{Fabiano:2023uhg}, adding some new observations.
In Section~\ref{sec:intro} and~\ref{sec:rhospace} we briefly summarized the main properties of the noncommutative spacetime of interest, $\varrho$-Minkowski. In Section~\ref{sec:QG} we described the properties of the quantum group $\mathcal{C}_\varrho(P)$, motivating our interest in bicrossproduct structures and expanding our previous work with an explicit construction of the bicrossproduct splitting of Lie algebra-type deformations of Poincar\'e coming from classical $r$-matrices. In Section~\ref{sec:QUEA}, then, we surveyed the main features of the dual QUEA, $U_\varrho(\mathfrak{p})$, explaining in which sense the bicrossproduct structure can be regarded as a generalisation of a semidirect product in the context of Hopf algebras.
Section~\ref{sec:star}, then, dealt with the problem of realising the algebra of noncommutative functions on the $\varrho$-Minkowski spacetime via the definition of different $\star$-products related to the coproduct structures we found in the previous section for the QUEA.
Section~\ref{sec:qa} discussed the issue coming from the possibility of having different bases for the QUEA and the need for a ``physical'' basis to which associate experimentable physical quantities. In this spirit, the $\varrho$-Poincar\'e QLA $\mathfrak{iso}_\varrho(1,3)$ was introduced and found to be connected to the QUEA in the bicrossproduct basis.

\section*{Acknowledgments}
We would like firstly to thank our collaborators Giuseppe Fabiano, Giulia Gubitosi, Fedele Lizzi and Patrizia Vitale, with whom we wrote the paper~\cite{Fabiano:2023uhg} that inspired this work.
We are then grateful to Paolo Aschieri and Andrzej Borowiec for fruitful and illuminating discussions.

We acknowledge financial support from the doctoral school of the University of Wrocław and the SONATA BIS grant 2021/42/E/ST2/00304 from the National Science Centre (NCN), Poland.

\bibliographystyle{JHEP}
\bibliography{references}
\end{document}